\documentstyle[seceq,preprint]{ptptex}

\newcommand{\bra}[1]{\langle {#1} |}     
\newcommand{\ket}[1]{| {#1} \rangle}     
\newcommand{\wtilde}[1]{\widetilde{#1}} 

\title{
Deformed Boson Scheme in Time-Dependent Variational Method.II
}
\subtitle{
Deformation of the su(2)- and the su(1,1)-Algebra in the 
Schwinger Boson Representation}    

\author{
Atsushi {\sc Kuriyama}, 
Constan\c{c}a {\sc Provid\^encia}$^{*}$, \\
Jo\~ao da {\sc Provid\^encia}$^{*}$, Yasuhiko {\sc Tsue}$^{**}$ 
and Masatoshi {\sc Yamamura}
}

\inst{
Faculty of Engineering, Kansai University, Suita 564-8680, Japan\\
$^{*}$Departamento de Fisica, Universidade de Coimbra, P-3000 Coimbra, 
Portugal\\
$^{**}$Physics Division, Faculty of Science, Kochi University, 
Kochi 780-8520, Japan
}

\recdate{
}

\abst{
In this paper as a continuation of Part I, the case of two 
kinds of boson operators is treated. The deformation of the 
coherent states for the $su(2)$- and the $su(1,1)$-algebra and 
their related deformed algebras are discussed in various 
form including the most popular form. 
}

\begin{document}

\maketitle

\section{Introduction}

Part (II), a continuation of (I),\cite{KPPTY} 
is concerned with the deformed boson scheme for the case of two 
kinds of boson operators. 
There exist two reasons why we intend to investigate this case. 
We mentioned in \S 6 of (I) that the deformed boson scheme 
displays its real ability in the present case. 
In the case of two kinds of boson operators, we know two spin 
systems which obey the $su(2)$- and $su(1,1)$-algebra 
in the Schwinger boson representations.\cite{S} 
As was reviewed in Ref.\citen{KPTYs}, we can construct the 
coherent states which are suitable for obtaining the classical 
counterparts of these two spin systems. 
These counterparts are useful for describing the time-evolution 
of these spin systems in the framework of the time-dependent 
variational method. 
For example, with the help of the $su(1,1)$-spin, the damped and 
amplified oscillation can be described in the 
conservative form.\cite{TKY} 
Therefore, as a natural course, it becomes interesting to investigate 
the deformation of the coherent states. 
This is the first reason. The second is related with the 
deformed algebras. Quantum mechanically, the $su(2)_q$- and 
the $su(1,1)_q$-algebra are quite interesting and various aspects of 
the $su(2)_q$-algebra have been investigated. 
Especially, the investigation based on the form 
$[x]_q=(q^x-q^{-x})/(q-q^{-1})$ is well known.\cite{Mac} 
The present authors also presented an idea how to 
derive the Holstein-Primakoff representation for the 
$su(2)_q$- and the $su(1,1)_q$-algebra\cite{PYK} 
in the framework of MYT boson mapping.\cite{MYT} 
Then, it is also a quite natural course to investigate the 
$su(2)_q$- and $su(1,1)_q$-algebra in our deformed 
boson scheme. The above two are main reasons why we present Part (II). 

In (I), the boson coherent state was deformed by a function 
$f(x)$ which determines $[x]_q$. Boson system treated in (II) 
consists of two kinds of boson operators and under a certain 
principle, the coherent states for the $su(2)$- and the 
$su(1,1)$-spin in the Schwinger boson representation are deformed by 
maximally three independent functions. 
Algebraically, it may be enough to take up two independent 
functions. Therefore, by changing the forms of these 
three functions, we are able to derive various types of the 
deformations including the most popular form for the 
$su(2)_q$-algebra. An interesting point is found in the fact 
that the Holstein-Primakoff boson representations for the 
$su(2)$- and the $su(1,1)$-spin are possible deformations 
of the $su(2)$- and the $su(1,1)$-algebra in the Schwinger 
representation. This form was already used for describing 
the damped and amplified oscillational motion in the 
$su(2)$-spin system.\cite{KPTY1} 

In \S 2, the $su(2)$- and the $su(1,1)$-spin system 
expressed in terms of two kinds of boson operators and their 
classical counterparts will be derived. 
The starting coherent states are of the same forms as those 
appearing in Ref.\citen{KPTYs}. 
Section 3 will be devoted in obtaining the deformed coherent 
states. Its basic idea is the use of three, algebraically, two 
independent functions. 
Further, the classical counterparts will be discussed. 
In \S 4, the $su(2)_q$- and the $su(1,1)_q$-algebra 
will be obtained in our deformed boson scheme and 
various forms will be discussed. In \S 5, the deformations 
named as the pseudo $su(2)$- and the pseudo $su(1,1)$-algebra will 
be discussed and the Holstein-Primakoff representation will be 
derived in the Schwinger representation by an appropriate 
select of the functions characterizing the deformation. 
Finally, in \S 6, some concluding remarks, together with 
the subsequent problem, will be given.

\section{The $su(2)$- and the $su(1,1)$-spin system and 
their classical counterparts}

In Part I, we investigated the deformed boson scheme 
in many-body system composed of one kind of boson 
operator $({\hat c}, {\hat c}^*)$. 
In the present paper (Part II), we will treat the case of 
many-body system consisting of two kinds of boson 
operators $({\hat a} , {\hat a}^*)$ and $({\hat b}, {\hat b}^*)$. 
The operators $({\hat a} , {\hat a}^*)$ and 
$({\hat b}, {\hat b}^*)$ obey the same relations as those of 
$({\hat c}, {\hat c}^*)$, which are 
shown in Eqs. (I$\cdot$2$\cdot$1)$\sim$(I$\cdot$2$\cdot$6).
For the present system, we know two spin systems, 
which obey the $su(2)$- and the $su(1,1)$-algebra. 
Hereafter, various relations for these systems will be 
shown in parallel form such as (Eq.a) and (Eq.b). 
The generators $({\hat S}_\pm^0, {\hat S}_0)$ for the 
$su(2)$-spin and $({\hat T}_\pm^0, {\hat T}_0)$ for the 
$su(1,1)$-spin are written down in the form  
\begin{subequations}\label{2-1}
\begin{eqnarray}
& &{\hat S}_+^0 = {\hat a}^*{\hat b} \ , \qquad
{\hat S}_-^0 = {\hat b}^*{\hat a} \ , \qquad
{\hat S}_0 = ({\hat a}^*{\hat a}-{\hat b}^*{\hat b})/2 \ , 
\label{2-1a}\\
& &{\hat T}_+^0 = {\hat a}^*{\hat b}^* \ , \qquad
{\hat T}_-^0 = {\hat b}{\hat a} \ , \qquad
{\hat T}_0 = ({\hat a}^*{\hat a}+{\hat b}{\hat b}^*)/2 \ . 
\label{2-1b}
\end{eqnarray}
\end{subequations}
Further, the following operators ${\hat S}$ and ${\hat T}$ are 
introduced in each spin system : 
\begin{subequations}\label{2-2}
\begin{eqnarray}
& &{\hat S} = ({\hat b}^*{\hat b}+{\hat a}^*{\hat a})/2 \ , 
\label{2-2a}\\
& &{\hat T} = ({\hat b}{\hat b}^*-{\hat a}^*{\hat a})/2 \ . 
\label{2-2b}
\end{eqnarray}
\end{subequations}
The commutation relations are given in the form 
\begin{subequations}\label{2-3}
\begin{eqnarray}
& &[{\hat S}_+^0 , {\hat S}_-^0 ]=+ 2{\hat S}_0 \ , \qquad
[{\hat S}_0 , {\hat S}_\pm^0 ] = \pm {\hat S}_{\pm}^0 \ , \qquad
[{\hat S}_{\pm,0}^0 , {\hat S} ] = 0 \ , 
\label{2-3a}\\
& &[{\hat T}_+^0 , {\hat T}_-^0 ]= - 2{\hat T}_0 \ , \qquad
[{\hat T}_0 , {\hat T}_\pm^0 ] = \pm {\hat T}_{\pm}^0 \ , \qquad
[{\hat T}_{\pm,0}^0 , {\hat T} ] = 0 \ . 
\label{2-3b}
\end{eqnarray}
\end{subequations}

In some papers by the present authors, we investigated two forms 
of normalized wave packets, which are expressed in the 
following forms : 
\begin{subequations}\label{2-4}
\begin{eqnarray}
\ket{c_+^0}&=& \left(\sqrt{\Gamma_0}\right)^{-1}
 \exp (\gamma{\hat S}_+^0)\exp (\delta{\hat b}^*) \ket{0} 
 \nonumber\\
&=& \left(\sqrt{\Gamma_0}\right)^{-1}
 \exp (\gamma{\hat a}^*{\hat b})\exp (\delta{\hat b}^*) \ket{0} \ ,
  \nonumber\\
\Gamma_0&=&\exp (|\delta|^2(1+|\gamma|^2)) \ , 
\label{2-4a}\\
\ket{c_-^0}&=& \left(\sqrt{\Gamma_0}\right)^{-1}
 \exp (\gamma{\hat T}_+^0)\exp (\delta{\hat b}^*) \ket{0} 
 \nonumber\\
&=& \left(\sqrt{\Gamma_0}\right)^{-1}
 \exp (\gamma{\hat a}^*{\hat b}^*)\exp (\delta{\hat b}^*) \ket{0} \ ,
  \nonumber\\
\Gamma_0&=&(1-|\gamma|^2)^{-1}\exp (|\delta|^2/(1-|\gamma|^2)) \ . 
\label{2-4b}
\end{eqnarray}
\end{subequations}
The plus ($+$) and the minus ($-$) indicate the wave packets 
for the $su(2)$- and the $su(1,1)$-spin system, respectively. 
The quantities $\gamma$ and $\delta$ denote complex parameters 
and our aim is to investigate the behaviors of these parameters. 
The forms (\ref{2-4}) show that the states $\ket{c_{\pm}^0}$ 
are coherently superposed in terms of various states obtained by 
operating ${\hat S}_+^0$ and ${\hat T}_+^0$ successively on 
the state which is a linear combination of the states 
with the minimum weights 
$({\hat S}_-^0\exp(\delta{\hat b}^*)\ket{0}
={\hat T}_-^0\exp(\delta{\hat b}^*)\ket{0}=0)$. 
For the state (\ref{2-4a}) and (\ref{2-4b}), 
we can prove the following relation : 
\begin{subequations}\label{2-5}
\begin{equation}
{\hat \gamma}^0\ket{c_+^0}
=\gamma(1-\epsilon({\hat N}_b+\epsilon)^{-1})\ket{c_+^0} \ , 
\qquad
{\hat \delta}^0\ket{c_+^0}=\delta\ket{c_+^0}
\label{2-5a}
\end{equation}
for (\ref{2-4a}) and 
\begin{equation}
{\hat \gamma}^0\ket{c_-^0}
=\gamma\ket{c_-^0} \ , 
\qquad
{\hat \delta}^0\ket{c_-^0}=\delta\ket{c_-^0}
\label{2-5b}
\end{equation}
\end{subequations}
for (\ref{2-4b}), where ${\hat \gamma}^0$ and ${\hat \delta}^0$ 
are defined as 
\begin{subequations}\label{2-6}
\begin{equation}
{\hat \gamma}^0={\hat S}_-^0({\hat N}_b+1+\epsilon)^{-1} \ , 
\qquad
{\hat \delta}^0={\hat b}
\label{2-6a}
\end{equation}
for (\ref{2-4a}) and 
\begin{equation}
{\hat \gamma}^0=({\hat N}_b+1+\epsilon)^{-1}{\hat T}_-^0 \ , 
\qquad
{\hat \delta}^0=[1-{\hat N}_a({\hat N}_b+1+\epsilon)^{-1}]{\hat b}
\label{2-6b}
\end{equation}
\end{subequations}
for (\ref{2-4b}), respectively. 
Of course, ${\hat N}_a$ and ${\hat N}_b$ denote the boson 
number operators : 
\begin{equation}\label{2-7}
{\hat N}_a={\hat a}^*{\hat a} \ , \qquad
{\hat N}_b={\hat b}^*{\hat b} \ .
\end{equation}
The symbol $\epsilon$ denotes an infinitesimal parameter and 
$\epsilon({\hat N}_b+\epsilon)^{-1}$ in Eq.(\ref{2-5a}) plays 
a role of the projection operator for the states 
$\{\ket{n}=(\sqrt{n!})^{-1}({\hat b}^*)^n\ket{0} ; 
n=0,1,2,\cdots \}$ : 
\begin{equation}\label{2-8}
\epsilon({\hat N}_b+\epsilon)^{-1}\ket{n}
=\delta_{n,0}\ket{n} \ . \qquad (\epsilon\longrightarrow 0) 
\end{equation}
Then, for large value of $|\delta|$, we can regard 
$\ket{c_+^0}$ as 
$\ket{c_+^0}\sim (1-\epsilon({\hat N}_b+\epsilon)^{-1})\ket{c_+^0}$. 
Thus, the states (\ref{2-4a}) and (\ref{2-4b}) are regarded as the 
eigenstates of ${\hat \gamma}^0$ and ${\hat \delta}^0$ defined 
in the relations (\ref{2-6a}) and (\ref{2-6b}), respectively. 

The time-dependent variational method for the states 
$\ket{c_\pm^0}$ starts with the following relation :
\begin{equation}\label{2-9}
\delta\int \bra{c_{\pm}^0} i\partial_t - {\hat H} \ket{c_\pm^0}
dt=0 \ .
\end{equation}
The states $\ket{c_\pm^0}$ satisfy the relation 
\begin{equation}\label{2-10}
\bra{c_\pm^0}i\partial_t \ket{c_\pm^0}
=\frac{i}{2}(\gamma^*{\dot \gamma}-{\dot \gamma}^*\gamma)
\frac{\partial \Gamma_0}{\partial |\gamma|^2}\cdot \Gamma_0^{-1}
+\frac{i}{2}(\delta^*{\dot \delta}-{\dot \delta}^*\delta)
\frac{\partial \Gamma_0}{\partial |\delta|^2}\cdot \Gamma_0^{-1} \ .
\end{equation}
Then, let us define the following quantities : 
\begin{equation}\label{2-11}
x=\gamma\sqrt{(\partial \Gamma_0/\partial |\gamma|^2)\cdot \Gamma_0^{-1}}
\ , \qquad
y=\delta\sqrt{(\partial \Gamma_0/\partial |\delta|^2)\cdot 
\Gamma_0^{-1}} \ .
\end{equation}
With the use of the new parameters $x$ and $y$, the relation 
(\ref{2-10}) is rewritten as 
\begin{equation}\label{2-12}
\bra{c_\pm^0} i\partial_t \ket{c_\pm^0}
=(i/2)\cdot(x^*{\dot x}-{\dot x}^*x)
+(i/2)\cdot(y^*{\dot y}-{\dot y}^*y) \ .
\end{equation}
The above is called the canonicity condition and in the 
sense of the time-dependent variational method, 
$(x, x^*)$ and $(y, y^*)$ can be regarded as the boson-type 
canonical variables in classical mechanics. 
The relation (\ref{2-11}) tells us that $x$ and $y$ can be 
expressed in terms of $\gamma$, $\delta$, $|\gamma|^2$ and 
$|\delta|^2$. In the present case, with the use of the forms 
(\ref{2-4a}) and (\ref{2-4b}) for $\Gamma_0$, 
$\gamma$ and $\delta$ can be expressed inversely 
in terms of $x$, $y$, $|x|^2$ and $|y|^2$. 
For the $su(2)$-spin system, we have 
\begin{subequations}\label{2-13}
\begin{equation}\label{2-13a}
\gamma=x\left(\sqrt{|y|^2-|x|^2}\right)^{-1} \ , \qquad
\delta=y\sqrt{(|y|^2-|x|^2)\cdot |y|^{-2}} \ .
\end{equation}
In the case of the $su(1,1)$-spin, we have 
\begin{equation}\label{2-13b}
\gamma=x\left(\sqrt{|y|^2+1+|x|^2}\right)^{-1} \ , \qquad
\delta=y\sqrt{|y|^2+1}\cdot\left(\sqrt{|y|^2+1+|x|^2}\right)^{-1} \ .
\end{equation}
\end{subequations}

Next, we investigate the classical counterpart of the operators 
${\hat \gamma}^0$ and ${\hat \delta}^0$. 
First, we note the relation 
\begin{equation}\label{2-14}
\bra{c_\pm^0}{\hat \gamma}^0\ket{c_\pm^0}=\gamma \ , \qquad
\bra{c_\pm^0}{\hat \delta}^0\ket{c_\pm^0}=\delta \ .
\end{equation}
From the argument given in the relation (\ref{2-8}), the 
former of the relation (\ref{2-14}) is approximated. 
Further, we define the difference of any function 
$F({\hat N}_a, {\hat N}_b)$ in the form 
\begin{eqnarray}\label{2-15}
& &\Delta_{{\hat N}_a}^{(\pm)} F({\hat N}_a, {\hat N}_b)
=\pm \left[ F({\hat N}_a\pm 1, {\hat N}_b)
-F({\hat N}_a, {\hat N}_b)\right] \ , \nonumber\\
& &\Delta_{{\hat N}_b}^{(\pm)} F({\hat N}_a, {\hat N}_b)
=\pm \left[ F({\hat N}_a, {\hat N}_b\pm 1)
-F({\hat N}_a, {\hat N}_b)\right] \ .
\end{eqnarray}
With the use of the difference (\ref{2-15}), the commutation 
relations of $({\hat \gamma}^0, {\hat \gamma}^{0*})$ and 
$({\hat \delta}^0, {\hat \delta}^{0*})$ for the $su(2)$-spin 
system are given in the form 
\begin{subequations}\label{2-16}
\begin{eqnarray}\label{2-16a}
& &[{\hat \gamma}^0 , {\hat \gamma}^{0*} ]
=(\Delta_{{\hat N}_a}^{(+)}-\Delta_{{\hat N}_b}^{(-)}
-\Delta_{{\hat N}_a}^{(+)}\Delta_{{\hat N}_b}^{(-)})
({\hat \gamma}^{0*}{\hat \gamma}^0) \ , \nonumber\\
& &[{\hat \delta}^0 , {\hat \delta}^{0*} ]
=\Delta_{{\hat N}_b}^{(+)}
({\hat \delta}^{0*}{\hat \delta}^0) \ , \nonumber\\
& &[{\hat \gamma}^0 , {\hat \delta}^{0} ]
={\hat \delta}^0{\hat \gamma}^0\cdot \epsilon 
\left[1-\epsilon({\hat N}_b+1+\epsilon)^{-1}\right]^{-1}\cdot 
\Delta_{{\hat N}_b}^{(+)}({\hat N}_b+\epsilon)^{-1} \ , \nonumber\\
& &[{\hat \gamma}^0 , {\hat \delta}^{0*} ]
={\hat \delta}^{0*}{\hat \gamma}^0
\left[(1-\epsilon({\hat N}_b+1+\epsilon)^{-1})^{-1}
({\hat \delta}^0{\hat \delta}^{0*})\right] \nonumber\\
& &\qquad\qquad\qquad\qquad
\times 
\Delta_{{\hat N}_b}^{(+)}\left[
(1-\epsilon({\hat N}_b+1+\epsilon)^{-1})
({\hat \delta}^0{\hat \delta}^{0*})^{-1}\right] \ .
\end{eqnarray}
For the case of the $su(1,1)$-spin system, we have 
\begin{eqnarray}\label{2-16b}
& &[{\hat \gamma}^0 , {\hat \gamma}^{0*} ]
=(\Delta_{{\hat N}_a}^{(+)}+\Delta_{{\hat N}_b}^{(+)}
+\Delta_{{\hat N}_a}^{(+)}\Delta_{{\hat N}_b}^{(+)})
({\hat \gamma}^{0*}{\hat \gamma}^0) \ , \nonumber\\
& &[{\hat \delta}^0 , {\hat \delta}^{0*} ]
=\Delta_{{\hat N}_b}^{(+)}
({\hat \delta}^{0*}{\hat \delta}^0) \ , \nonumber\\
& &[{\hat \gamma}^0 , {\hat \delta}^{0} ]=0 \ , \nonumber\\
& &[{\hat \gamma}^0 , {\hat \delta}^{0*} ]
={\hat \gamma}^0{\hat \delta}^{0*}
\left[(1-\Delta_{{\hat N}_a}^{(-)})
({\hat \gamma}^0{\hat \gamma}^{0*})\right]^{-1}\cdot 
\Delta_{{\hat N}_b}^{(-)}\left[
(1-\Delta_{{\hat N}_a}^{(-)})
({\hat \gamma}^0{\hat \gamma}^{0*})\right] \ . \ \ \ \ \ \ 
\end{eqnarray}
\end{subequations}
The relation (\ref{2-16a}) contains the infinitesimal 
parameter $\epsilon$. After operating this relation 
on any state, we should make the limitation 
$\epsilon \rightarrow 0$. 
The relations (\ref{2-13a}) and (\ref{2-13b}) tell us that 
$\gamma$ and $\delta$ are expressed in terms of the 
canonical variables $(x, x^*)$ and $(y, y^*)$. 
Then, we can calculate the Poisson bracket for these 
variables defined in the form 
\begin{equation}\label{2-17}
[ A, B]_P=(\partial_x A\cdot \partial_{x^*}B
-\partial_{x^*}A\cdot \partial_x B)
+(\partial_y A\cdot \partial_{y^*}B
-\partial_{y^*}A\cdot \partial_y B) \ .
\end{equation}
For the $su(2)$-spin system, the result is as follows : 
\begin{subequations}\label{2-18}
\begin{eqnarray}\label{2-18a}
& &[ \gamma , \gamma^* ]_P = (\partial_{N_a}-\partial_{N_b})|\gamma|^2 \ , 
\nonumber\\
& &[ \delta, \delta^* ]_P=\partial_{N_b}|\delta|^2  \ , \nonumber\\
& &[ \gamma, \delta ]_P = 0 \ , \nonumber\\
& &[ \gamma, \delta^* ]_P=\delta^*\gamma\cdot |\delta|^2
\partial_{N_b}|\delta|^{-2} \ .
\end{eqnarray}
For the $su(1,1)$-spin system, we have 
\begin{eqnarray}\label{2-18b}
& &[ \gamma , \gamma^* ]_P = (\partial_{N_a}+\partial_{N_b})|\gamma|^2 \ , 
\nonumber\\
& &[ \delta, \delta^* ]_P=\partial_{N_b}|\delta|^2  \ , \nonumber\\
& &[ \gamma, \delta ]_P = 0 \ , \nonumber\\
& &[ \gamma, \delta^* ]_P=\gamma\delta^*\cdot |\gamma|^{-2}
\partial_{N_b}|\gamma|^{2} \ .
\end{eqnarray}
\end{subequations}
Here, $\partial_{N_a}$ and $\partial_{N_b}$ denote the 
differential with respect to $N_a$ and $N_b$, respectively. 
The variables $N_a$ and $N_b$ are the expectation values 
of ${\hat N}_a$ and ${\hat N}_b$, respectively : 
\begin{eqnarray}\label{2-19}
N_a=\bra{c_\pm^0}{\hat N}_a\ket{c_\pm^0}
&=&|\gamma|^2\partial \Gamma_0/\partial |\gamma|^2 \cdot 
\Gamma_0^{-1}=|x|^2 \ , \nonumber\\
N_b=\bra{c_\pm^0}{\hat N}_b\ket{c_\pm^0}
&=&|\delta|^2\partial\Gamma_0/\partial|\delta|^2\cdot 
\Gamma_0^{-1}\mp|\gamma|^2
\partial\Gamma_0/\partial|\gamma|^2 \cdot \Gamma_0^{-1} \nonumber\\
&=&|y|^2\mp |x|^2 \ .
\end{eqnarray}
For the above derivation, the following formula is 
useful : 
\begin{equation}\label{2-20}
|\gamma|^2\partial_{N_a}|\delta|^2
\mp |\gamma|^2\partial_{N_b}|\delta|^2
-|\delta|^2\partial_{N_b}|\gamma|^2=0 \ .
\end{equation}
For the commutation relations (\ref{2-16a}) and (\ref{2-16b}), 
we perform the replacement 
\begin{equation}\label{2-21}
[\ , \ ] \longrightarrow [ \ , \ ]_P \ , \qquad
\Delta \longrightarrow \partial \ .
\end{equation}
Then, at the limits $\epsilon \rightarrow 0$ and 
$\partial_{N_a}\partial_{N_b}|\gamma|^2 \rightarrow 0$, 
the relations (\ref{2-16}) are reduced to the relations (\ref{2-18}). 
The limit $\partial_{N_a}\partial_{N_b}|\gamma|^2 \rightarrow 0$ 
means the neglect of quantal fluctuation around 
$(\partial_{N_a}+\partial_{N_b})|\gamma|^2$. 
Thus, for $\gamma$ and $\delta$ given in the relation (\ref{2-13}), 
we have the following correspondence : 
\begin{equation}\label{2-22}
{\hat \gamma}^0 \sim \gamma \ , \qquad
{\hat \delta}^0 \sim \delta \ .
\end{equation}
The above means that $\gamma$ and $\delta$ introduced as the 
variational parameters are classical counterparts of the 
operators ${\hat \gamma}^0$ and ${\hat \delta}^0$, respectively. 

The expectation values of $({\hat S}_{\pm,0}^0, {\hat S})$ 
and $({\hat T}_{\pm,0}^0, {\hat T})$ for the states 
$\ket{c_\pm^0}$ are expressed in the form 
\begin{subequations}\label{2-23}
\begin{eqnarray}
& &\bra{c_+^0}{\hat S}_+^0\ket{c_+^0}
=\gamma^*|\delta|^2=x^*\sqrt{2s-|x|^2} \ , \nonumber\\
& &\bra{c_+^0}{\hat S}_-^0\ket{c_+^0}
=\gamma|\delta|^2=x\sqrt{2s-|x|^2} \ , \nonumber\\
& &\bra{c_+^0}{\hat S}_0\ket{c_+^0}
=-(1-|\gamma|^2)|\delta|^2/2=|x|^2-s \ , \nonumber\\
& &\bra{c_+^0}{\hat S}\ket{c_+^0}
=(1+|\gamma|^2)|\delta|^2/2=s \ , 
\label{2-23a}\\
& &\bra{c_-^0}{\hat T}_+^0\ket{c_-^0}
=\gamma^*(1-|\gamma|^2+|\delta|^2)(1-|\gamma|^2)^{-2}
=x^*\sqrt{2t+|x|^2} \ , \nonumber\\
& &\bra{c_-^0}{\hat T}_-^0\ket{c_-^0}
=\gamma(1-|\gamma|^2+|\delta|^2)(1-|\gamma|^2)^{-2}
=x\sqrt{2t+|x|^2} \ , \nonumber\\
& &\bra{c_-^0}{\hat T}_0\ket{c_-^0}
=(1+|\gamma|^2)(1-|\gamma|^2+|\delta|^2)(1-|\gamma|^2)^{-2}/2
=|x|^2+t \ , \nonumber\\
& &\bra{c_-^0}{\hat T}\ket{c_-^0}
=(1-|\gamma|^2+|\delta|^2)(1-|\gamma|^2)^{-1}/2=(|y|^2+1)/2=t \ . 
\label{2-23b}
\end{eqnarray}
\end{subequations}
The above forms are nothing but classical forms of the 
$su(2)$- and the $su(1,1)$-generators in the 
Holstein-Primakoff boson representation. The quantities $s$ 
and $t$ shown in the relations (\ref{2-23a}) and (\ref{2-23b}) 
denote the magnitudes of the $su(2)$- and the $su(1,1)$-spin, 
respectively, in the classical sense.

\section{Deformation of the coherent state $\ket{c_\pm^0}$}

We make deformation for the state $\ket{c_{\pm}^0}$ by 
introducing three functions ${\wtilde f}$, ${\wtilde g}$ 
and ${\wtilde h}$ which play the same role as that of 
${\wtilde f}$ appearing in the state (I$\cdot$2$\cdot$8). 
Our idea for the deformation starts with the following form : 
\begin{subequations}\label{3-1}
\begin{eqnarray}
& &\ket{c_+}=\left(\sqrt{\Gamma}\right)^{-1}
\exp\left(\gamma{\hat a}^*{\wtilde f}({\hat N}_a)\cdot
{\wtilde g}({\hat N}_b){\hat b}\right)
\cdot\exp\left(\delta{\hat b}^*{\wtilde h}({\hat N}_b)^{-1}\right)
\ket{0} \ , 
\label{3-1a}\\
& &\ket{c_-}=\left(\sqrt{\Gamma}\right)^{-1}
\exp\left(\gamma{\hat a}^*{\wtilde f}({\hat N}_a)\cdot
{\hat b}^*{\wtilde g}({\hat N}_b)\right)
\cdot\exp\left(\delta{\hat b}^*{\wtilde h}({\hat N}_b)^{+1}\right)
\ket{0} \ . 
\label{3-1b}
\end{eqnarray}
\end{subequations}
We can see in the expressions (\ref{3-1a}) and (\ref{3-1b}) 
that two parts of the exponential forms in the states 
(\ref{2-4a}) and (\ref{2-4b}) are deformed by 
${\wtilde f}({\hat N}_a)$, ${\wtilde g}({\hat N}_b)$ and 
${\wtilde h}({\hat N}_b)$. 
The quantity $\Gamma$ denotes the normalization constant. 
The states $\ket{c_\pm}$ can be rewritten in the form 
\begin{eqnarray}\label{3-2}
\ket{c_\pm}=\left(\sqrt{\Gamma}\right)^{-1}
\sum_{m,n}&\!\!&
\frac{\gamma^m\delta^n}{\sqrt{m!n!}}
\left(\sqrt{\frac{n!}{(n\mp m)!}}\right)^{\pm 1}
f(m)g(n\mp m)^{\mp 1}(g(n)\cdot h(n)^{-1})^{\pm 1} \nonumber\\
& &\times \left(\sqrt{m!(n\mp m)!}\right)^{-1}({\hat a}^*)^m
({\hat b}^*)^{n\mp m} \ket{0} \ .
\end{eqnarray}
Here, $f(k)$, $g(k)$ and $h(k)$ are defined through the relation 
\begin{eqnarray}\label{3-3}
& &{\wtilde f}(k)=f(k+1)f(k)^{-1} \ , \quad
{\wtilde g}(k)=g(k+1)g(k)^{-1} \ , \quad
{\wtilde h}(k)=h(k+1)h(k)^{-1} \ . \nonumber\\
& &
\qquad\qquad\qquad\qquad\qquad\qquad\qquad\qquad\qquad\qquad
 (k=0, 1, 2, \cdots)
\end{eqnarray}
The normalization constant $\Gamma$ is obtained as 
\begin{equation}\label{3-4}
\Gamma=\sum_{m,n}\frac{(|\gamma|^2)^m(|\delta|^2)^n}{m!n!}
\left(\frac{n!}{(n\mp m)!}\right)^{\pm 1}
f(m)^2g(n\mp m)^{-2}
(g(n)\cdot h(n)^{-1})^{\pm 2} \ .
\end{equation}
The state (\ref{3-2}) can be rewritten as 
\begin{eqnarray}\label{3-5}
& &\ket{c_\pm}=\sqrt{\Gamma_0/\Gamma}\cdot \Omega_\pm ({\hat N}_a, 
{\hat N}_b)\ket{c_\pm^0} \ , \nonumber\\
& &\Omega_\pm({\hat N}_a, {\hat N}_b)=
f({\hat N}_a)g({\hat N}_b)^{\mp 1}
\left(g({\hat N}_b\pm{\hat N}_a)\cdot h({\hat N}_b\pm{\hat N}_a)^{-1}
\right)^{\pm 1} \ .
\end{eqnarray}
Concerning the functions $f(k)$, $g(k)$ and $h(k)$ introduced in 
the expression (\ref{3-2}), we must give a remark. 
They are well-behaved and obey 
\begin{equation}\label{3-6}
f(k) \ , \quad g(k) \ , \quad h(k) \ > \ 0 \ . \qquad
(k=0, 1, 2, \cdots)
\end{equation}
Further, they obey the condition 
\begin{equation}\label{3-7}
f(0)=h(0)=h(1)=1 \ , \qquad f(1)g(1)=g(0) \ .
\end{equation}
The above is derived through the following process : 
The states $\ket{c_\pm}$ can be expressed in the form 
\begin{subequations}\label{3-8}
\begin{eqnarray}
\sqrt{\Gamma}\ket{c_+}&=&
(1+{\wtilde h}(0)^{-1}\delta{\hat b}^*
+{\wtilde f}(0){\wtilde g}(0){\wtilde h}(0)^{-1}\gamma\delta{\hat a}^*
+\cdots )\ket{0} \nonumber\\
&=&
(f(0)h(0)^{-1}+f(0)h(1)^{-1}\delta{\hat b}^*
+f(1)g(0)^{-1}g(1)h(1)^{-1}\gamma\delta{\hat a}^*
+\cdots )\ket{0} \ , \nonumber\\
& &\label{3-8a}\\
\sqrt{\Gamma}\ket{c_-}&=&
(1+{\wtilde h}(0)\delta{\hat b}^*
+{\wtilde f}(0){\wtilde g}(0)\gamma{\hat a}^*{\hat b}^*
+\cdots )\ket{0} \nonumber\\
&=&
(f(0)h(0)+f(0)h(1)\delta{\hat b}^*
+f(1)g(0)^{-1}g(1)h(0)\gamma{\hat a}^*{\hat b}^*
+\cdots )\ket{0} \ . \nonumber\\
& &\label{3-8b}
\end{eqnarray}
\end{subequations}
For the coefficients of ${\hat b}^*\ket{0}$ and 
${\hat a}^*\ket{0}$ in the state (\ref{3-8a}), it is 
permitted to set up the condition 
\begin{subequations}\label{3-9}
\begin{eqnarray}\label{3-9a}
& &f(0)h(0)^{-1}=1 \ , \qquad f(0)h(1)^{-1}={\wtilde h}(0)^{-1}=1 \ , 
\nonumber\\
& &f(1)g(0)^{-1}g(1)h(1)^{-1}={\wtilde f}(0){\wtilde g}(0)
{\wtilde h}(0)^{-1}=1 \ .
\end{eqnarray}
In the same way, the state (\ref{3-8b}) gives us 
\begin{eqnarray}\label{3-9b}
& &f(0)h(0)=1 \ , \qquad f(0)h(1)={\wtilde h}(0)=1 \ , 
\nonumber\\
& &f(1)g(0)^{-1}g(1)h(0)={\wtilde f}(0){\wtilde g}(0)
=1 \ .
\end{eqnarray}
\end{subequations}
From the relations (\ref{3-9a}) and (\ref{3-9b}), we have 
the condition (\ref{3-7}). 
As was shown in the above, the deformation is performed by, maximally, 
three independent functions $f(x)$, $g(x)$ and $h(x)$. 
However, we pay an attention to the case mentioned below. 
As is clear from the relations (\ref{2-2}) and (\ref{2-3}), 
the terms $g({\hat N}_b\pm {\hat N}_a)\cdot h({\hat N}_b\pm{\hat N}_a)^{-1}$ 
in $\Omega_\pm({\hat N}_a, {\hat N}_b)$ appearing in the 
relation (\ref{3-5}) commute with all the generators of the 
$su(2)$- and the $su(1,1)$-algebra, respectively. 
Then, for the generators of the algebras, the terms 
$g({\hat N}_b\pm{\hat N}_a)\cdot h({\hat N}_b\pm {\hat N}_a)^{-1}$ 
do not give any influence and, from this reason, we restrict 
ourselves to the case 
\begin{eqnarray}
& &g({\hat N}_b\pm{\hat N}_a)\cdot h({\hat N}_b\pm{\hat N}_a)^{-1}
=1 \ , \label{3-10}\\
\hbox{\rm i.e.,} \qquad\quad & &
\Omega_\pm({\hat N}_a,{\hat N}_b)=f({\hat N}_a)\cdot 
g({\hat N}_b)^{\mp 1} \ .
\label{3-11} 
\end{eqnarray}
This means that the deformation of $\ket{c_\pm^0}$ is 
characterized by two functions $f(x)$ and $g(x)$.

Next, we define the operators ${\hat \gamma}$ and ${\hat \delta}$ 
in the form 
\begin{eqnarray}
& &{\hat \gamma}=\Omega_{\pm}({\hat N}_a, {\hat N}_b) 
{\hat \gamma}^0 \Omega_{\pm}({\hat N}_a, {\hat N}_b)^{-1} \ , 
\label{3-12}\\
& &{\hat \delta}=\Omega_{\pm}({\hat N}_a, {\hat N}_b) 
{\hat \delta}^0 \Omega_{\pm}({\hat N}_a, {\hat N}_b)^{-1} \ . 
\label{3-13}
\end{eqnarray}
With the use of the state (\ref{3-5}), the relations (\ref{2-5a}) 
and (\ref{2-5b}) give us 
\begin{subequations}\label{3-14}
\begin{eqnarray}
& &{\hat \gamma}\ket{c_+}
=\gamma(1-\epsilon({\hat N}_b+\epsilon)^{-1})\ket{c_+} \ ,
\qquad
{\hat \delta}\ket{c_+}=\delta\ket{c_+} \ , 
\label{3-14a}\\
& &{\hat \gamma}\ket{c_-}=\gamma\ket{c_-} \ , \qquad
{\hat \delta}\ket{c_-}=\delta\ket{c_-} \ .
\label{3-14b}
\end{eqnarray}
\end{subequations}
The commutation relations for ${\hat \gamma}$, ${\hat \gamma}^*$, 
${\hat \delta}$ and ${\hat \delta}^*$ for the $su(2)$-spin 
system are given as 
\begin{subequations}\label{3-15}
\begin{eqnarray}\label{3-15a}
& &[{\hat \gamma} , {\hat \gamma}^* ]
=\left( \Delta_{{\hat N}_a}^{(+)}-\Delta_{{\hat N}_b}^{(-)}
-\Delta_{{\hat N}_a}^{(+)}\Delta_{{\hat N}_b}^{(-)}\right)
({\hat \gamma}^*{\hat \gamma}) \ , \nonumber\\
& &[ {\hat \delta} , {\hat \delta}^* ]
=\Delta_{{\hat N}_b}^{(+)}({\hat \delta}^*{\hat \delta}) \ ,
\nonumber\\
& &[ {\hat \gamma} , {\hat \delta} ] 
={\hat \delta}{\hat \gamma}\cdot \epsilon 
\left[1-\epsilon({\hat N}_b+1+\epsilon)^{-1}\right]^{-1}
\cdot \Delta_{{\hat N}_b}^{(+)}({\hat N}_b+\epsilon)^{-1} \ ,
\nonumber\\
& &[ {\hat \gamma} , {\hat \delta}^* ] 
={\hat \delta}^*{\hat \gamma}\cdot 
\left[(1-\epsilon({\hat N}_b+1+\epsilon)^{-1})^{-1}
({\hat \delta}{\hat \delta}^*)\right] \nonumber\\
& &\qquad\qquad\qquad\times 
\Delta_{{\hat N}_b}^{(+)}
\left[(1-\epsilon({\hat N}_b+1+\epsilon)^{-1})
({\hat \delta}{\hat \delta}^*)^{-1}\right] \ .
\end{eqnarray}
In the case of the $su(1,1)$-spin system, we have
\begin{eqnarray}\label{3-15b}
& &[{\hat \gamma} , {\hat \gamma}^* ]
=\left( \Delta_{{\hat N}_a}^{(+)}+\Delta_{{\hat N}_b}^{(+)}
+\Delta_{{\hat N}_a}^{(+)}\Delta_{{\hat N}_b}^{(+)}\right)
({\hat \gamma}^*{\hat \gamma}) \ , \nonumber\\
& &[ {\hat \delta} , {\hat \delta}^* ]
=\Delta_{{\hat N}_b}^{(+)}({\hat \delta}^*{\hat \delta}) \ ,
\nonumber\\
& &[ {\hat \gamma} , {\hat \delta} ] 
=0 \ ,
\nonumber\\
& &[ {\hat \gamma} , {\hat \delta}^* ] 
={\hat \gamma}{\hat \delta}^*\cdot
\left[(1-\Delta_{{\hat N}_a}^{(-)})({\hat \gamma}{\hat \gamma}^*)
\right]^{-1}
\cdot \Delta_{{\hat N}_b}^{(-)}
\left[(1-\Delta_{{\hat N}_a}^{(-)})
({\hat \gamma}{\hat \gamma}^*)\right] \ .
\end{eqnarray}
\end{subequations}
The relation (\ref{3-15a}) contains the infinitesimal 
parameter $\epsilon$. After operating this relation on any 
state, we should make the limitation $\epsilon \rightarrow 0$. 
It may be interesting to see that the commutation relations 
(\ref{3-15a}) and (\ref{3-15b}) are of the forms 
quite similar to those of the relations (\ref{2-16a}) 
and (\ref{2-16b}).

Now, let us investigate the classical counterpart of the 
present case. In the same meaning as that in the case of 
$\ket{c_{\pm}^0}$, the parameters $\gamma$ and $\delta$ are the 
eigenvalue of the operators ${\hat \gamma}$ and ${\hat \delta}$, 
respectively. This fact suggests us that we can treat the deformed 
wave packet $\ket{c_\pm}$ in the same way as that for the 
state $\ket{c_\pm^0}$ presented in \S 2. 
The state $\ket{c_\pm}$ satisfies the same relation as 
that shown in Eq.(\ref{2-10}) : 
\begin{equation}\label{3-16}
\bra{c_\pm}i\partial_t \ket{c_\pm}
=(i/2)(\gamma^*{\dot \gamma}-{\dot \gamma}^*\gamma)
(\partial \Gamma/\partial |\gamma|^2)\cdot \Gamma^{-1}
+(i/2)(\delta^*{\dot \delta}-{\dot \delta}^*\delta)
(\partial \Gamma/\partial |\delta|^2)\cdot \Gamma^{-1} \ .
\end{equation}
The relation (\ref{3-16}) can be rewritten as 
\begin{equation}\label{3-17}
\bra{c_\pm}i\partial_t \ket{c_\pm}
=(i/2)(x^*{\dot x}-{\dot x}^* x)
+(i/2)(y^*{\dot y}-{\dot y}^* y) \ .
\end{equation}
Here, $x$ and $y$ are defined as 
\begin{equation}\label{3-18}
x=\gamma\sqrt{(\partial\Gamma/\partial |\gamma|^2)\cdot \Gamma^{-1}} \ , 
\qquad
y=\delta\sqrt{(\partial\Gamma/\partial |\delta|^2)\cdot \Gamma^{-1}} \ . 
\end{equation}
New parameters $(x, x^*)$ and $(y, y^*)$ can be regarded as 
canonical variables and, in principle, $\gamma$ and $\delta$ 
can be expressed in terms of those canonical variables 
by solving inversely Eq.(\ref{3-18}). 
However, in general, the explicit solving is impossible. 
In the case of $\ket{c_\pm^0}$, it is possible. 
The expectation values of ${\hat N}_a$ and ${\hat N}_b$ for 
$\ket{c_\pm}$ are given in the form 
\begin{eqnarray}\label{3-19}
N_a=\bra{c_\pm}{\hat N}_a \ket{c_\pm}
&=&|\gamma|^2 \partial\Gamma/\partial |\gamma|^2\cdot \Gamma^{-1}
=|x|^2 \ , \nonumber\\
N_b=\bra{c_\pm}{\hat N}_b \ket{c_\pm}
&=&|\delta|^2 \partial\Gamma/\partial |\delta|^2\cdot \Gamma^{-1}
\mp |\gamma|^2 \partial \Gamma/\partial |\gamma|^2\cdot \Gamma^{-1}
\nonumber\\
&=&|y|^2\mp |x|^2 \ .
\end{eqnarray}
Here, we used the relation (\ref{3-18}). The above are the same 
as those given in the relation (\ref{2-19}). 
Therefore, we can prove that $(\gamma, \gamma^*)$ and $(\delta, \delta^*)$ 
in the present case obey the same relations as those for 
$(\gamma, \gamma^*)$ and $(\delta, \delta^*)$ in the case 
treated in \S 2. 
For example, the Poisson brackets for $(\gamma, \gamma^*)$ and 
$(\delta, \delta^*)$ in the present case are given in the same forms 
as those shown in the relation (\ref{2-18a}) and (\ref{2-18b}). 
Thus, under the replacement (\ref{2-21}), the commutation 
relations (\ref{3-15a}) and (\ref{3-15b}) are reduced 
to the relations (\ref{2-18a}) and (\ref{2-18b}). 
Of course, the terms related with the quantal fluctuation 
are neglected. This is in the same situation as that in \S 2. 
Then, we have the correspondence 
\begin{equation}\label{3-20}
{\hat \gamma}\sim \gamma \ , \qquad 
{\hat \delta} \sim \delta \ .
\end{equation}
In this way, we obtained the classical counterpart of 
the deformed boson scheme in parallel with the case of 
the conventional boson coherent state $\ket{c_\pm^0}$.

\section{The $su(2)_q$- and the $su(1,1)_q$-algebra in the present 
deformed boson scheme}

It may be an interesting problem to investigate the 
$su(2)_q$- and $su(1,1)_q$-algebra in the present deformed 
boson scheme. Conventionally, the $su(2)_q$-algebra is formulated 
in terms of $({\hat S}_\pm , {\hat S}_0, [2{\hat S}_0]_q)$, 
which obey the following commutation relation : 
\begin{subequations}\label{4-1}
\begin{equation}\label{4-1a}
[{\hat S}_0 , {\hat S}_\pm ]=\pm {\hat S}_\pm \ , \qquad
[{\hat S}_+ , {\hat S}_- ] = + [2{\hat S}_0]_q \ .
\end{equation}
In analogy with the above case, the $su(1,1)_q$-algebra is 
formulated by setting up the following relation 
for the set $({\hat T}_\pm , {\hat T}_0 , [2{\hat T}_0]_q)$ : 
\begin{equation}\label{4-1b}
[{\hat T}_0 , {\hat T}_\pm ]=\pm {\hat T}_\pm \ , \qquad
[{\hat T}_+ , {\hat T}_- ] = - [2{\hat T}_0]_q \ .
\end{equation}
\end{subequations}
For the operators $[2{\hat S}_0]_q$ and $[2{\hat T}_0]_q$, 
conventionally, 
\begin{equation}\label{4-2}
[x]_q=(q^x-q^{-x})/(q-q^{-1}) \ . \qquad (q\ :\ \hbox{\rm real})
\end{equation}
We investigate the above algebras in terms of the space 
composed of two kinds of the boson operators.

For the above-mentioned aim, let us define the operators 
\begin{eqnarray}\label{4-3}
& &{\hat \alpha}=\Omega_\pm({\hat N}_a, {\hat N}_b){\hat a}
\Omega_{\pm}({\hat N}_a, {\hat N}_b)^{-1} \ , \nonumber\\
& &{\hat \beta}=\Omega_\pm({\hat N}_a, {\hat N}_b)^{\mp 1}{\hat b}
\Omega_{\pm}({\hat N}_a, {\hat N}_b)^{\pm 1} \ .
\end{eqnarray}
Here, $\Omega_\pm({\hat N}_a, {\hat N}_b)$ is defined in 
Eq.(\ref{3-11}). Then, for ${\hat S}_\pm$ and ${\hat T}_\pm$, 
we give the forms 
\begin{subequations}\label{4-4}
\begin{eqnarray}
& &{\hat S}_+={\hat \alpha}^*{\hat \beta} \ , \qquad
{\hat S}_-={\hat \beta}^*{\hat \alpha} \ , 
\label{4-4a}\\
& &{\hat T}_+={\hat \alpha}^*{\hat \beta}^* \ , \qquad
{\hat T}_-={\hat \beta}{\hat \alpha} \ . 
\label{4-4b}
\end{eqnarray}
\end{subequations}
With the use of the operators (\ref{4-3}), ${\hat S}_\pm$ and 
${\hat T}_\pm$ can be expressed as follows : 
\begin{subequations}\label{4-5}
\begin{eqnarray}
& &{\hat S}_+=f({\hat N}_a)^{-1}g({\hat N}_b)
{\hat S}_+^0 f({\hat N}_a)g({\hat N}_b)^{-1} \ , \qquad
{\hat S}_+^0={\hat a}^*{\hat b} \ , \nonumber\\
& &{\hat S}_-=f({\hat N}_a)g({\hat N}_b)^{-1}
{\hat S}_-^0 f({\hat N}_a)^{-1}g({\hat N}_b) \ , \qquad
{\hat S}_-^0={\hat b}^*{\hat a} \ , 
\label{4-5a}\\
& &{\hat T}_+=f({\hat N}_a)^{-1}g({\hat N}_b)^{-1}
{\hat T}_+^0 f({\hat N}_a)g({\hat N}_b) \ , \qquad
{\hat T}_+^0={\hat a}^*{\hat b}^* \ , \nonumber\\
& &{\hat T}_-=f({\hat N}_a)g({\hat N}_b)
{\hat T}_-^0 f({\hat N}_a)^{-1}g({\hat N}_b)^{-1} \ , \qquad
{\hat T}_-^0={\hat b}{\hat a} \ . 
\label{4-5b}
\end{eqnarray}
\end{subequations}
Following the conventional manner, further, we define 
$[2{\hat S}_0]_q$ and $[2{\hat T}_0]_q$ through the relations 
\begin{subequations}\label{4-6}
\begin{eqnarray}
& &[ {\hat S}_+ , {\hat S}_- ]= + [2{\hat S}_0]_q \ , 
\label{4-6a}\\
& &[ {\hat T}_+ , {\hat T}_- ]= - [2{\hat T}_0]_q \ . 
\label{4-6b}
\end{eqnarray}
\end{subequations}
The definitions (\ref{4-6a}) and (\ref{4-6b}) give us the 
operators $[2{\hat S}_0]_q$ and $[2{\hat T}_0]_q$ in the form 
\begin{subequations}\label{4-7}
\begin{eqnarray}
& &[2{\hat S}_0]_q=[{\hat N}_a]_f[{\hat N}_b+1]_g
-[{\hat N}_a+1]_f[{\hat N}_b]_g \ , 
\label{4-7a}\\
& &[2{\hat T}_0]_q=[{\hat N}_a+1]_f[{\hat N}_b+1]_g
-[{\hat N}_a]_f[{\hat N}_b]_g \ . 
\label{4-7b}
\end{eqnarray}
\end{subequations}
Here, $[x]_f$ and $[x]_g$ are given as 
\begin{equation}\label{4-8}
[x]_f=xf(x)^{-2}f(x-1)^2 \ , \qquad
[x]_g=xg(x)^{-2}g(x-1)^2 \ .
\end{equation}
The above form was given in the relation (I$\cdot$2$\cdot$19). 
The operators ${\hat S}_0$ and ${\hat T}_0$ given in the relations 
(\ref{2-1a}) and (\ref{2-1b}) give us 
\begin{subequations}\label{4-9}
\begin{eqnarray}
& &[ {\hat S}_0 , {\hat S}_\pm ]=\pm {\hat S}_\pm \ , 
\label{4-9a}\\
& &[ {\hat T}_0 , {\hat T}_\pm ]=\pm {\hat T}_\pm \ . 
\label{4-9b}
\end{eqnarray}
\end{subequations}
The above argument suggests us that $({\hat S}_\pm , [2{\hat S}_0]_q)$ 
and $({\hat T}_\pm , [2{\hat T}_0]_q)$ defined in the relations 
(\ref{4-5a}), (\ref{4-7a}), (\ref{4-5b}) and (\ref{4-7b}) 
form the $su(2)_q$- and the $su(1,1)_q$-algebra, 
respectively. 
The operators $({\hat S}_\pm , [2{\hat S}_0]_q)$ may be 
functions of ${\hat S}$, which commutes with them. 
The case of $({\hat T}_\pm , [2{\hat T}_0]_q)$ is also 
in the same situation as that in the above : 
They may be functions of ${\hat T}$, which 
commutes with them. The forms (\ref{4-5a}), (\ref{4-7a}), 
(\ref{4-5b}) and (\ref{4-7b}) can be rewritten as follows : 
\begin{subequations}\label{4-10}
\begin{eqnarray}
& &{\hat S}_+={\hat E}_a^*{\hat E}_b 
\sqrt{[{\hat S}+{\hat S}_0+1]_f [{\hat S}-{\hat S}_0]_g}
=\sqrt{[{\hat S}+{\hat S}_0]_f [{\hat S}-{\hat S}_0+1]_g}
{\hat E}_a^*{\hat E}_b \ , \nonumber\\
& &{\hat S}_-={\hat E}_b^*{\hat E}_a 
\sqrt{[{\hat S}+{\hat S}_0]_f [{\hat S}-{\hat S}_0+1]_g}
=\sqrt{[{\hat S}+{\hat S}_0+1]_f [{\hat S}-{\hat S}_0]_g}
{\hat E}_b^*{\hat E}_a \ , \nonumber\\
& &[2{\hat S}_0]_q
=[{\hat S}+{\hat S}_0]_f [{\hat S}-{\hat S}_0+1]_g
-[{\hat S}+{\hat S}_0+1]_f [{\hat S}-{\hat S}_0]_g \ , 
\label{4-10a}\\
& &{\hat T}_+={\hat E}_a^*{\hat E}_b^* 
\sqrt{[{\hat T}_0-{\hat T}+1]_f [{\hat T}_0+{\hat T}]_g}
=\sqrt{[{\hat T}_0-{\hat T}]_f [{\hat T}_0+{\hat T}-1]_g}
{\hat E}_a^*{\hat E}_b^* \ , \nonumber\\
& &{\hat T}_-={\hat E}_b{\hat E}_a 
\sqrt{[{\hat T}_0-{\hat T}]_f [{\hat T}_0+{\hat T}-1]_g}
=\sqrt{[{\hat T}_0-{\hat T}+1]_f [{\hat T}_0+{\hat T}]_g}
{\hat E}_b{\hat E}_a \ , \nonumber\\
& &[2{\hat T}_0]_q
=[{\hat T}_0-{\hat T}+1]_f [{\hat T}_0+{\hat T}]_g
-[{\hat T}_0-{\hat T}]_f [{\hat T}_0+{\hat T}-1]_g \ . 
\label{4-10b}
\end{eqnarray}
\end{subequations}
Here, ${\hat S}$, ${\hat T}$, ${\hat S}_0$ and ${\hat T}_0$ are 
given in the relations (\ref{2-2}) and (\ref{2-1}), respectively. 
The operator $({\hat E}_c, {\hat E}_c^*)$ for 
$c=a, b$ are defined as 
\begin{equation}\label{4-11}
{\hat E}_c=\left(\sqrt{{\hat N}_c+1}\right)^{-1}\ {\hat c} \ , 
\qquad
{\hat E}_c^*={\hat c}^*\ 
\left(\sqrt{{\hat N}_c+1}\right)^{-1} \ . 
\qquad ({\hat N}_c={\hat c}^*{\hat c})
\end{equation}
The property is as follows : 
\begin{equation}\label{4-12}
{\hat E}_c{\hat E}_c^*=1 \ , \qquad
{\hat N}_c{\hat E}_c^*{\hat E}_c
={\hat E}_c^*{\hat E}_c{\hat N}_c={\hat N}_c \ .
\end{equation}

In (I), we showed some examples for the deformed boson. 
In this section, three of them will be applied for 
the $su(2)_q$- and the $su(1,1)_q$-algebra. \\
(i) The most popular form :
\begin{eqnarray}\label{4-13}
& &f(n)=\sqrt{n(q-q^{-1})/(q^n-q^{-n})}\ f(n-1) \ , 
\nonumber\\
& &g(n)=\sqrt{n(q-q^{-1})/(q^n-q^{-n})}\ g(n-1) \ . 
\end{eqnarray}
In this case, both the $su(2)_q$- and the $su(1,1)_q$-generators 
are expressed in the form 
\begin{subequations}\label{4-14}
\begin{eqnarray}
& &{\hat S}_+^{\rm (i)}
=\frac{1}{\sqrt{{\hat S}\!-\!{\hat S}_0\!+\!1}}{\hat S}_+^0
\frac{1}{\sqrt{{\hat S}\!+\!{\hat S}_0\!+\!1}}
\sqrt{\frac{q^{({\hat S}-{\hat S}_0)}-q^{-({\hat S}-{\hat S}_0)}}
{q-q^{-1}}}
\sqrt{\frac{q^{({\hat S}+{\hat S}_0+1)}-q^{-({\hat S}+{\hat S}_0+1)}}
{q-q^{-1}}} \ , \nonumber\\
& &{\hat S}_-^{\rm (i)}
=
\sqrt{\frac{q^{({\hat S}-{\hat S}_0)}-q^{-({\hat S}-{\hat S}_0)}}
{q-q^{-1}}}
\sqrt{\frac{q^{({\hat S}+{\hat S}_0+1)}-q^{-({\hat S}+{\hat S}_0+1)}}
{q-q^{-1}}} 
\frac{1}{\sqrt{{\hat S}\!+\!{\hat S}_0\!+\!1}}{\hat S}_-^0
\frac{1}{\sqrt{{\hat S}\!-\!{\hat S}_0\!+\!1}}
\ , \nonumber\\
& &[2{\hat S}_0]_q^{\rm (i)}=\frac{q^{2{\hat S}_0}-q^{-2{\hat S}_0}}
{q-q^{-1}} \ , 
\label{4-14a}\\
& &{\hat T}_+^{\rm (i)}
={\hat T}_+^0 \frac{1}{\sqrt{({\hat T}_0+{\hat T})
({\hat T}_0-{\hat T}+1)}}
\sqrt{\frac{q^{({\hat T}_0+{\hat T})}-q^{-({\hat T}_0+{\hat T})}}
{q-q^{-1}}}
\sqrt{\frac{q^{({\hat T}_0-{\hat T}+1)}-q^{-({\hat T}_0-{\hat T}+1)}}
{q-q^{-1}}} \ , \nonumber\\
& &{\hat T}_-^{\rm (i)}
=
\sqrt{\frac{q^{({\hat T}_0+{\hat T})}-q^{-({\hat T}_0+{\hat T})}}
{q-q^{-1}}}
\sqrt{\frac{q^{({\hat T}_0-{\hat T}+1)}-q^{-({\hat T}_0-{\hat T}+1)}}
{q-q^{-1}}} 
\frac{1}{\sqrt{({\hat T}_0+{\hat T})({\hat T}_0-{\hat T}+1)}}
{\hat T}_-^0 \ , \nonumber\\
& &[2{\hat T}_0]_q^{\rm (i)}=\frac{q^{2{\hat T}_0}-q^{-2{\hat T}_0}}
{q-q^{-1}} \ , 
\label{4-14b}
\end{eqnarray}
\end{subequations}
As is clear from the form of $[2{\hat S}_0]_q$ shown in 
Eq.(\ref{4-14a}), the functions (\ref{4-13}) gives us the 
most popular form for the $su(2)_q$- and the $su(1,1)_q$-algebra. \\
(ii)$_{\rm a}$ 
The form presented by Penson and Solomon\cite{Penson} 
for the $su(2)_q$-algebra : 
\begin{subequations}
\begin{equation}
f(n)=q^{-(n-1)/2} f(n-1) \ , \qquad
g(n)=q^{-(n-1)/2} g(n-1) \ .
\label{4-15a}
\end{equation}
\end{subequations}
In this case, we have 
\begin{subequations}
\begin{eqnarray}\label{4-16a}
& &{\hat S}_+^{\rm (ii)}=q^{{\hat S}-1/2}\cdot {\hat S}_+^0 \ , \qquad
{\hat S}_-^{\rm (ii)}
=q^{{\hat S}-1/2}\cdot {\hat S}_-^0 \ , \nonumber\\
& &[2{\hat S}_0]_q^{\rm (ii)}=q^{2({\hat S}-1/2)}\cdot 2{\hat S}_0 \ .
\end{eqnarray}
\end{subequations}
\setcounter{equation}{14}
(ii)$_{\rm b}$ 
The form presented by Penson and Solomon 
for the $su(1,1)_q$-algebra : 
\begin{subequations}
\setcounter{equation}{14\stepcounter{equation}}
\begin{equation}
f(n)=q^{+(n-1)/2} f(n-1) \ , \qquad
g(n)=q^{-(n-1)/2} g(n-1) \ .
\label{4-15b}
\end{equation}
\end{subequations}
In this case, we have 
\begin{subequations}
\setcounter{equation}{15}
\begin{eqnarray}\label{4-16b}
& &{\hat T}_+^{\rm (ii)}=q^{{\hat T}-1/2}\cdot {\hat T}_+^0 \ , \qquad
{\hat T}_-^{\rm (ii)}=q^{{\hat T}-1/2}\cdot {\hat T}_-^0 \ , \nonumber\\
& &[2{\hat T}_0]_q^{\rm (ii)}=q^{2({\hat T}-1/2)}\cdot 2{\hat T}_0 \ .
\end{eqnarray}
\end{subequations}
(iii)$_{\rm a}$ 
Modified form for the $su(2)_q$-algebra : 
\begin{subequations}
\begin{eqnarray}
& &f(n)=\sqrt{n(1-q^{2})/(1-q^{2n})} f(n-1) \ , \nonumber\\
& &g(n)=\sqrt{n(1-q^{2})/(1-q^{2n})} g(n-1) \ .
\label{4-17a}
\end{eqnarray}
\end{subequations}
In this case, we have the form 
\begin{subequations}
\begin{eqnarray}\label{4-18a}
& &{\hat S}_+^{\rm (iii)}=q^{{\hat S}-1/2}\cdot {\hat S}_+^{\rm (i)} \ , \qquad
{\hat S}_-^{\rm (iii)}
=q^{{\hat S}-1/2}\cdot {\hat S}_-^{\rm (i)} \ , \nonumber\\
& &[2{\hat S}_0]_q^{\rm (iii)}
=q^{2({\hat S}-1/2)}\cdot [2{\hat S}_0]_q^{\rm (i)} 
\ .
\end{eqnarray}
\end{subequations}
\setcounter{equation}{16}
(iii)$_{\rm b}$ 
Modified form for the $su(1,1)_q$-algebra : 
\begin{subequations}
\setcounter{equation}{16}
\setcounter{equation}{16\refstepcounter{equation}}
\begin{eqnarray}
& &f(n)=\sqrt{n(1-q^{-2})/(1-q^{-2n})} f(n-1) \ , \nonumber\\
& &g(n)=\sqrt{n(1-q^2)/(1-q^{2n})} g(n-1) \ .
\label{4-17b}
\end{eqnarray}
\end{subequations}
In this case, we have the form 
\begin{subequations}
\setcounter{equation}{17}
\begin{eqnarray}\label{4-18b}
& &{\hat T}_+^{\rm (iii)}
=q^{{\hat T}-1/2}\cdot {\hat T}_+^{\rm (i)} \ , \qquad
{\hat T}_-^{\rm (iii)}
=q^{{\hat T}-1/2}\cdot {\hat T}_-^{\rm (i)} \ , \nonumber\\
& &[2{\hat T}_0]_q^{\rm (iii)}
=q^{2({\hat T}-1/2)}\cdot [2{\hat T}_0]_q^{\rm (i)} 
\ .
\end{eqnarray}
\end{subequations}
The above form is proposed by the present authors in (I) and 
it may be interesting to see that the form is in the 
intermediate situation between the forms (i) and (ii).

\section{The pseudo $su(2)$- and the pseudo $su(1,1)$-algebra}

With the aim of describing a boson system interacting 
with external field, the present authors proposed three 
forms of coherent states in the $su(2)$-spin system. 
In this section, we reinvestigate these forms from 
the viewpoint of the deformed boson scheme presented in \S 4. 
For this purpose, let us consider the following case : 
\begin{equation}\label{5-1}
f(n)=\left(\sqrt{1+q(n-1)}\right)^{-1} f(n-1)\ , \quad
g(n)=\sqrt{1+r(n-1)} g(n-1)\ .
\end{equation}
Here, $q$ and $r$ denote real parameters. It should be 
noted that, as was shown in \S 4, there exist two functions 
$f(n)$ and $g(n)$ which can be chosen independently. 
Then, ${\hat S}_\pm$ and ${\hat T}_\pm$ given in the relations 
(\ref{4-5a}) and (\ref{4-5b}), respectively, are written down 
in the form 
\begin{subequations}\label{5-2}
\begin{eqnarray}
& &{\hat S}_+={\hat a}^*\sqrt{1+q{\hat N}_a}
\left(\sqrt{1+r{\hat N}_b}\right)^{-1}\ {\hat b} \ ,
\nonumber\\
& &{\hat S}_-={\hat b}^*\left(\sqrt{1+r{\hat N}_b}\right)^{-1}
\sqrt{1+q{\hat N}_a}\ {\hat a} \ ,
\label{5-2a}\\
& &{\hat T}_+={\hat a}^*\sqrt{1+q{\hat N}_a}\ 
{\hat b}^* \left(\sqrt{1+r{\hat N}_b}\right)^{-1} \ ,
\nonumber\\
& &{\hat T}_-=\left(\sqrt{1+r{\hat N}_b}\right)^{-1}\ 
{\hat b} \sqrt{1+q{\hat N}_a}\ {\hat a} \ .
\label{5-2b}
\end{eqnarray}
\end{subequations}
Through the definitions (\ref{4-6a}) and (\ref{4-6b}), 
$[2{\hat S}_0]_q$ and $[2{\hat T}_0]_q$ can be expressed as 
follows : 
\begin{subequations}\label{5-3}
\begin{eqnarray}
& &[2{\hat S}_0]_q=+2({\hat S}_0)_q+2({\hat S}_0)_p\cdot
\epsilon ({\hat N}_b+\epsilon)^{-1} \ , 
\label{5-3a}\\
& &[2{\hat T}_0]_q=-2({\hat T}_0)_q-2({\hat T}_0)_p\cdot
\epsilon ({\hat N}_b+\epsilon)^{-1} \ , 
\label{5-3b}
\end{eqnarray}
\end{subequations}
\begin{subequations}\label{5-4}
\begin{eqnarray}
({\hat S}_0)_q&=&
(1-r)/2\cdot\left[
{\hat N}_a-{\hat N}_b-q{\hat N}_a(1-{\hat N}_a+2{\hat N}_b)\right]
-r/2\cdot \left[1+2q{\hat N}_a\right] \nonumber\\
& &-r(1-r)/2\cdot \big[
(1+2q{\hat N}_a){\hat N}_b^2 (1+r{\hat N}_b)^{-1} \nonumber\\
& &-(1+{\hat N}_a)(1+q{\hat N}_a)(1+(r-2){\hat N}_b
-r{\hat N}_b^2)(1+r{\hat N}_b)^{-1}(1-r+r{\hat N}_b)^{-1}
\big] \ , \nonumber\\
& &\label{5-4a}\\
({\hat T}_0)_q&=&
(1-r)/2\cdot\left[
({\hat N}_a+{\hat N}_b+1)+q{\hat N}_a(1+{\hat N}_a+2{\hat N}_b)\right]
+r/2\cdot \left[1+2q{\hat N}_a\right] \nonumber\\
& &-r(1-r)/2\cdot \big[
(1+2q{\hat N}_a){\hat N}_b^2 (1+r{\hat N}_b)^{-1} \nonumber\\
& &-{\hat N}_a(1-q+q{\hat N}_a)(1+(r-2){\hat N}_b
-r{\hat N}_b^2)(1+r{\hat N}_b)^{-1}(1-r+r{\hat N}_b)^{-1}
\big] \ , \nonumber\\
& &\label{5-4b}
\end{eqnarray}
\end{subequations}
\begin{subequations}\label{5-5}
\begin{eqnarray}
({\hat S}_0)_p&=&
r/2\cdot (1+{\hat N}_a)(1+q{\hat N}_a) \nonumber\\
& &\times \left[
1-(1-r)(1+(r-2){\hat N}_b-r{\hat N}_b^2)(1+r{\hat N}_b)^{-1}
(1-r+r{\hat N}_b)^{-1}\right] \ , \nonumber\\
& &\label{5-5a}\\
({\hat T}_0)_p&=&
r/2\cdot {\hat N}_a(1-q+q{\hat N}_a) \nonumber\\
& &\times \left[
1-(1-r)(1+(r-2){\hat N}_b-r{\hat N}_b^2)(1+r{\hat N}_b)^{-1}
(1-r+r{\hat N}_b)^{-1}\right] \ . \nonumber\\
& &\label{5-5b}
\end{eqnarray}
\end{subequations}
Here, $\epsilon({\hat N}_b+\epsilon)^{-1}$ was already 
introduced in the relation (\ref{2-5a}) with its 
property (\ref{2-8}). For the relations (\ref{5-2a}) 
and (\ref{5-3a}), three cases $(q=0, r=0)$, $(q=0, r=1)$ 
and $(q>0 , r=1)$ were investigated in Ref.\citen{KPTY1}. 
The first is nothing but the Schwinger boson representation and 
the third enables us to describe a boson system interacting 
with the external field. 

Under the above-mentioned background, let us treat the cases 
$(q<0, r=1)$, $(q=0, r=1)$ and $(q>0, r=1)$ more systematically 
than that in Ref.\citen{KPTY1}. First, we define the following 
operators :
\begin{subequations}\label{5-6}
\begin{eqnarray}
& &{\wtilde c}={\hat E}_b^*{\hat a} \ , \qquad
{\wtilde c}^*={\hat a}^*{\hat E}_b \ , 
\label{5-6a}\\
& &{\wtilde c}={\hat E}_b{\hat a} \ , \qquad
{\wtilde c}^*={\hat a}^*{\hat E}_b^* \ . 
\label{5-6b}
\end{eqnarray}
\end{subequations}
Here, $({\hat E}_b , {\hat E}_b^*)$ is defined in the 
relation (\ref{4-11}). The property of the operator 
(\ref{5-6a}) is given by 
\begin{subequations}\label{5-7}
\begin{eqnarray}
& &[ {\wtilde c} , {\wtilde c}^* ]
=1-(1+{\hat N}_a)\cdot \epsilon({\hat N}_b+\epsilon)^{-1} \ ,
\nonumber\\
& &{\wtilde c}^*{\wtilde c}={\wtilde N}={\hat N}_a \ .
\label{5-7a}
\end{eqnarray}
For the operator (\ref{5-6a}), we have 
\begin{eqnarray}
& &[ {\wtilde c} , {\wtilde c}^* ]
=1+{\hat N}_a\cdot \epsilon({\hat N}_b+\epsilon)^{-1} \ ,
\nonumber\\
& &{\wtilde c}^*{\wtilde c}={\wtilde N}
={\hat N}_a-{\hat N}_a\cdot \epsilon({\hat N}_b+\epsilon)^{-1} \ .
\label{5-7b}
\end{eqnarray}
\end{subequations}
Then, if we restrict ourselves to the subspace which does 
not include the vacuum $\ket{0}$ and any state consisting 
only of the ${\hat a}$-boson, the operators 
$({\wtilde c} , {\wtilde c}^*)$ defined in the relations 
(\ref{5-6a}) and (\ref{5-6b}) can be regarded as boson operators 
and ${\wtilde N}$ denotes the boson number operator : 
\begin{equation}\label{5-8}
[ {\wtilde c} , {\wtilde c}^* ]=1 \ , \qquad
{\wtilde c}^*{\wtilde c}={\wtilde N} \ (={\hat N}_a) \ .
\end{equation}
Hereafter, we treat the above subspace.

With the use of the operators $({\wtilde c} , {\wtilde c}^*)$ 
and ${\wtilde N}$, we express ${\hat S}_\pm$ in the form 
(\ref{5-2a}) and $({\hat S}_0)_q$ in the form (\ref{5-4a}). 
Also, in the case of ${\hat T}_\pm$ in the form (\ref{5-2b}) 
and $({\hat T}_0)_q$ in the form (\ref{5-4b}), we have the 
following expressions : \\
(i)${}_{\rm a}$ $q<0$ : 
\begin{subequations}\label{5-9}
\begin{eqnarray}
& &|q|^{-1/2}{\hat S}_+={\wtilde c}^*
\sqrt{|q|^{-1}-{\wtilde N}} \ , \qquad
|q|^{-1/2}{\hat S}_-=\sqrt{|q|^{-1}-{\wtilde N}}\ {\wtilde c} \ , 
\nonumber\\
& &|q|^{-1}({\hat S}_0)_q={\wtilde N}-|q|^{-1}/2 \ , 
\label{5-9a}
\end{eqnarray}
(i)${}_{\rm b}$ $q<0$ : 
\begin{eqnarray}
& &|q|^{-1/2}{\hat T}_+={\wtilde c}^*
\sqrt{|q|^{-1}-{\wtilde N}} \ , \qquad
|q|^{-1/2}{\hat T}_-=\sqrt{|q|^{-1}-{\wtilde N}}\ {\wtilde c} \ , 
\nonumber\\
& &-|q|^{-1}({\hat T}_0)_q={\wtilde N}-|q|^{-1}/2 \ , 
\label{5-9b}
\end{eqnarray}
\end{subequations}
(ii)${}_{\rm a}$ $q=0$ : 
\begin{subequations}\label{5-10}
\begin{equation}
{\hat S}_+={\wtilde c}^* \ , \qquad {\hat S}_-={\wtilde c} \ ,
\qquad -2({\hat S}_0)_q=1 \ , 
\label{5-10a}
\end{equation}
(ii)${}_{\rm b}$ $q=0$ : 
\begin{equation}
{\hat T}_+={\wtilde c}^* \ , \qquad {\hat T}_-={\wtilde c} \ ,
\qquad 2({\hat T}_0)_q=1 \ , 
\label{5-10b}
\end{equation}
\end{subequations}
(iii)${}_{\rm a}$ $q>0$ : 
\begin{subequations}\label{5-11}
\begin{eqnarray}
& &q^{-1/2}{\hat S}_+={\wtilde c}^*
\sqrt{q^{-1}+{\wtilde N}} \ , \qquad
q^{-1/2}{\hat S}_-=\sqrt{q^{-1}+{\wtilde N}}\ {\wtilde c} \ , 
\nonumber\\
& &-q^{-1}({\hat S}_0)_q={\wtilde N}+q^{-1}/2 \ , 
\label{5-11a}
\end{eqnarray}
(iii)${}_{\rm b}$ $q>0$ : 
\begin{eqnarray}
& &q^{-1/2}{\hat T}_+={\wtilde c}^*
\sqrt{q^{-1}+{\wtilde N}} \ , \qquad
q^{-1/2}{\hat T}_-=\sqrt{q^{-1}+{\wtilde N}}\ {\wtilde c} \ , 
\nonumber\\
& &q^{-1}({\hat T}_0)_q={\wtilde N}+q^{-1}/2 \ . 
\label{5-11b}
\end{eqnarray}
\end{subequations}
We can see that $(|q|^{-1/2}{\hat S}_\pm , |q|^{-1}({\hat S}_0)_q)$ 
and $(|q|^{-1/2}{\hat T}_\pm , -|q|^{-1}({\hat T}_0)_q)$ form, respectively, 
the $su(2)$-algebras in the Holstein-Primakoff representation 
if $({\wtilde c} , {\wtilde c}^*)$ can be regarded as the boson 
operator strictly. 
However, as was already mentioned, $({\wtilde c} , {\wtilde c}^*)$ 
can be regarded as the boson operator in a certain subspace. 
In this sense, it may be permitted to call the above algebra 
the pseudo $su(2)$-algebra. Further, the sets 
$(q^{-1/2}{\hat S}_\pm , -q^{-1}({\hat S}_0)_q)$ and 
$(q^{-1/2}{\hat T}_\pm , q^{-1}({\hat T}_0)_q)$ form, 
respectively, the $su(1,1)$-algebras in the 
Holstein-Primakoff representation if 
$({\wtilde c} , {\wtilde c}^*)$ can be regarded as the 
boson operator strictly. 
In this sense, in the same meaning as that of the $su(2)$-algebra, 
it may be permitted to call the above algebra the pseudo 
$su(1,1)$-algebra. The sets $({\hat S}_\pm)$ and 
$({\hat T}_\pm)$ in $q=0$, respectively, 
behave as the boson operator formally. 
The quantities $|q|^{-1}/2$ and $q^{-1}/2$ denote the 
magnitudes of the $su(2)$- and the $su(1,1)$-spin. 

We showed deformations of the $su(2)$- and the $su(1,1)$-algebra 
in the Schwinger boson representation for three forms. 
Then, let us investigate the states constructed by 
these deformations. First, we consider the cases given in 
the relations (\ref{5-9a}), (\ref{5-10a}) and (\ref{5-11a}). 
The condition ${\hat S}_-\ket{{\wtilde 0}}=0$ gives us 
\begin{equation}\label{5-12a}
\ket{\wtilde 0}=({\hat E}_b^*)^\Lambda \ket{0} \ . 
\qquad
(\Lambda=1,2,3,\cdots)
\end{equation}
Successive operation of ${\wtilde c}^*$ on the state 
$\ket{\wtilde 0}$ gives us the states in which the 
number of the ${\hat b}$-boson decreases and the number of 
the ${\hat a}$-boson increases. 
Then, in order to make the form (\ref{5-9a}) be the 
$su(2)$-spin in the Holstein-Primakoff representation 
in the subspace which does not include the vacuum $\ket{0}$ 
and any state consisting only of the ${\hat a}$-boson, 
the following relation should be set up : 
\begin{eqnarray}\label{5-13a}
|q|^{-1}=2\sigma \ , \qquad
& &\sigma=0, 1/2, 1, 3/2, \cdots, (\Lambda-1)/2 \ , \nonumber\\
& &\sigma_0=-\sigma, -\sigma+1 , \cdots , \sigma-1, \sigma \ .
\end{eqnarray}
In the case of the relation (\ref{5-10a}), the number of the 
operation of ${\wtilde c}^*$ on the state $\ket{\wtilde 0}$ 
is restricted to 
\begin{equation}\label{5-14a}
n=0, 1, 2, \cdots, (\Lambda-1) \ .
\end{equation}
In order to make the form (\ref{5-11a}) be the $su(1,1)$-algebra 
in the Holstein-Primakoff representation in the subspace, the 
following relation is necessary : 
\begin{eqnarray}\label{5-15a}
q^{-1}=2\tau \ ,\qquad
& &\tau= \hbox{\rm positive\ but\ arbitrary} \ , \nonumber\\
& &\tau_0=\tau , \tau+1 ,\cdots , \tau+(\Lambda-1)/2 \ .
\end{eqnarray}
The relation (\ref{5-15a}) shows that $\tau_0$ cannot run to 
the infinity and, then, strictly speaking, the form (\ref{5-11a}) 
does not compose the $su(1,1)$-algebra. 
However, if $\Lambda$ is sufficiently large, we can regard 
the form (\ref{5-11a}) as the $su(1,1)$-spin approximately and this fact 
enables us to describe the damped and amplified oscillation of a 
boson system interacting with the external field. 
Next, we consider the cases (\ref{5-9b}), (\ref{5-10b}) and 
(\ref{5-11b}). In this case, also, ${\hat T}_-\ket{\wtilde 0}=0$ 
gives us 
\begin{equation}\label{5-12b}
\ket{\wtilde 0}=({\hat E}_b^*)^\Lambda \ket{0} \ . 
\qquad
(\Lambda=1,2,3,\cdots)
\end{equation}
Successive operation of ${\wtilde c}^*$ on the state 
$\ket{\wtilde 0}$ gives us the states in which the 
numbers of the ${\hat a}$- and ${\hat b}$-boson increase. 
Then, we have the following relations : 
\begin{eqnarray}\label{5-13b}
|q|^{-1}=2\sigma \ , \qquad
& &\sigma=0, 1/2, 1, 3/2, \cdots \ , \nonumber\\
& &\sigma_0=-\sigma, -\sigma+1 , \cdots , \sigma-1, \sigma \ .
\end{eqnarray}
\begin{equation}\label{5-14b}
n=0, 1, 2, \cdots \ , \qquad\qquad\qquad\qquad\qquad\qquad
\end{equation}
\begin{eqnarray}\label{5-15b}
q^{-1}=2\tau \ ,\qquad
& &\tau= \hbox{\rm positive\ but\ arbitrary} \ , \nonumber\\
& &\tau_0=\tau , \tau+1 , \tau+2 , \cdots \ .
\end{eqnarray}
The $su(2)$-algebra is compact and the $su(1,1)$-algebra 
is non-compact. From this fact, the difference between 
the two cases appears.

\section{Concluding remarks}

Following the basic idea presented in (I), in this paper, 
we investigated the deformation of the system 
obeying the $su(2)$- and the $su(1,1)$-algebra. 
With the use of two independent functions $f(x)$ and $g(x)$, 
the deformation is performed. One of the interesting points 
presented in this paper may be to be shown that the 
$su(2)_q$-algebra in the most popular form is nothing but 
one type of the deformations. 
The case of the $su(1,1)_q$-algebra is also in the same situation 
as the above. For example, the deformation discussed in \S 5 
is interesting, because this type was already used by the present 
authors for describing the damped and amplified oscillational motion 
in the $su(2)$-spin system.\cite{KPTY1} 
However, this treatment does not enable us to describe 
statistically mixed state in the system 
discussed in Ref.\citen{KPTY1}. 
For this problem, we applied the $su(2,1)$-algebra 
in three kinds of boson operators.\cite{KPTY2} 
In Part (III), we will discuss this case in the deformed 
boson scheme given in (I).



\end{document}